\documentclass[aps,prl,twocolumn,superscriptaddress]{revtex4-1}
\usepackage{graphicx}
\usepackage{amsmath}
\usepackage{braket}
\usepackage{url}

\usepackage[colorlinks,
linkcolor=blue,
anchorcolor=blue,
urlcolor=blue,
citecolor=blue]{hyperref}

\begin{document}

%\preprint{}

\title{Description of proton-rich nuclei in the $A \sim 20$ region with the Gamow shell model}


\author{N. Michel}\email[]{nicolas.michel@impcas.ac.cn}
\affiliation{Institute of Modern Physics, Chinese Academy of Sciences, Lanzhou 730000, China}
\affiliation{School of Nuclear Science and Technology, University of Chinese Academy of Sciences, Beijing 100049, China}
\author{J.G. Li}
\affiliation{School of Physics,  and   State Key  Laboratory  of  Nuclear  Physics   and  Technology, Peking University, Beijing  100871, China}
\author{F.R. Xu}
\affiliation{School of Physics,  and   State Key  Laboratory  of  Nuclear  Physics   and  Technology, Peking University, Beijing  100871, China}
\author{W. Zuo}\email[]{zuowei@impcas.ac.cn}
\affiliation{Institute of Modern Physics, Chinese Academy of Sciences, Lanzhou 730000, China}
\affiliation{School of Nuclear Science and Technology, University of Chinese Academy of Sciences, Beijing 100049, China}

%\thanks{}


\date{\today}

\begin{abstract}
  The Gamow shell model is utilized to describe nuclear observables of the weakly bound and resonance isotonic states of $^{16}$O at proton drip-line.
  It is hereby shown that the presence of continuum coupling leads to complex Coulomb contributions in the spectrum of these isotones.
  The necessity to include the effects of three-body forces, either by a direct calculation or by adding an $A$-dependence to the nucleon-nucleon interaction,
  already noticed in other theoretical models, is pointed out.
  It is also demonstrated that our approach is predictive for reaction observables.
\end{abstract}

\pacs{}
%\keywords{}{}

\maketitle

\section{Introduction}

Accelerators of new generation, based on the use of radioactive ion beams, have allowed to study nuclei up to proton and neutron drip-lines \cite{TANIHATA2013215,Motobayashi2014,BLANK2008403,RevModPhys.84.567}.
The location of the neutron drip-line is not well known experimentally,
as the heaviest drip-line neutron-rich nuclei to have been synthesized in accelerators are $^{31}$F and $^{40}$Mg \cite{wang2017ame2016,PhysRevLett.109.202503,Baumann20071022,PhysRevLett.122.052501}. 
Conversely, the proton drip-line has been reached experimentally up to $Z \sim 90$ since already more than a decade \cite{Thoenessen2004}.
Nevertheless, due to the very large Coulomb barrier for large $Z$, it is difficult to generate drip-line nuclei beyond $Z = 90$ \cite{Thoenessen2004}.
Indeed, to our knowledge, the heaviest proton drip-line nucleus to have been synthesized in the recent years is $^{219}$Np, for which $Z = 93$ \cite{2018PhLB..777..212Y}.
Moreover, unbound nuclei close to the proton drip-line bear a very long lifetime, often of a few milliseconds \cite{BLANK2008403},
so that it is not always clear whether the proton drip-line has been crossed or not during an experiment \cite{Thoenessen2004}.
In fact, to our knowledge, the exact location of the proton drip-line is only known before aluminum, for which $Z = 13$ \cite{Thoenessen2004}.
The production of nuclei bearing a large number of protons is also challenging in the domain of superheavies, where elements whose charge number is larger than 110 have been produced \cite{RevModPhys.84.567}.
The heaviest element to date to have been synthesized in a accelerator is the $Z = 118$ element, called Oganesson from its discoverer Y.T.~Oganessian \cite{PhysRevC.74.044602}.

Due to the large confining Coulomb barrier at proton drip-line, proton-rich nuclei can be usually described by models devised for well-bound nuclei,
such as shell model based on a basis of harmonic oscillator states (HO-SM) \cite{PhysRevC.80.011306,PhysRevLett.110.022502,PhysRevC.99.034313}.
However, the proton emitters, usually found in the $A \sim 110-150$ region, which have a very small particle-emission width, demand the use of elaborate methods to calculate their widths precisely.
The use of the two-potential method in the spherical case \cite{PhysRevC.56.1762},
or coupled-channel models in the non-adiabatic approach in the axially deformed case \cite{PhysRevLett.84.4549,PhysRevC.62.054315}, have been seen to be very successful for that matter.
In fact, very light proton-rich nuclei can bear the same unusual properties found at neutron drip-line, in which the Coulomb interaction plays very little role. 
For example, the ground states of $^{5}$Li and $^7$B are very broad resonance states, as they bear a width of about 1 MeV. While being more narrow, the ground states of $^6$Be and $^8$C are unbound by about 100 keV.
Moreover, proton-rich nuclei at drip-lines can also sustain halos, albeit less frequently than neutron-rich nuclei \cite{Al-Khalili2004}.
Indeed, the ground states of $^8$B and $^{13}$N and the first excited state of $^{17}$F are one-proton halos, and that of $^{17}$Ne is a two-proton halo \cite{Al-Khalili2004}. 
Among the features which can develop only at proton drip-line, one can also cite diproton emission, which has been discovered one to two decades ago \cite{PhysRevLett.89.102501,PhysRevC.72.054315}.
The theoretical description of diproton emitters demands to rigorously treat the asymptotic region.
For this, the use of the shell model embedded in the continuum along with cluster approximation has been seen to be successful,
as it could provide with diproton emission widths close to experimental data \cite{ROTUREAU200613}.
Consequently, the variety of phenomena induced by the proximity of proton-emission threshold shows that one has to describe the asymptotic part of many-body wave functions precisely.

For proton-rich nuclei whose number of nucleons is close to $20$, the Coulomb barrier is already rather large, so that one can expect protons to be confined in general in the nuclear zone.
Nevertheless, proton-emission threshold is low therein, so that effects due to the proximity of the continuum also appear in nuclear states.
Consequently, both Coulomb and continuum effects compete in proton-rich nuclei, so that continuum coupling has to be included explicitly.
For this, the Gamow shell model (GSM), which incorporates both continuum coupling and inter-nucleon correlations,
is the tool of choice for this type of applications \cite{PhysRevLett.89.042502,PhysRevC.96.024308,0954-3899-36-1-013101}.
Indeed, GSM has become a predictive tool of experimental interest for structure and reaction observables of nuclei at drip-line,
as it has been used to analyze the $^{14}$O(p,p) elastic reaction cross section \cite{DeGrancey:2016bez}, $^{25}$O spectroscopy \cite{PhysRevC.96.054322},
and the newly discovered $^{20,21}$B isotopes at neutron drip-line \cite{PhysRevLett.121.262502}.
GSM has also been introduced successfully to the use of realistic Hamiltonians, in a no-core approach \cite{PhysRevC.88.044318}, or in a core + valence nucleons picture,
using an effective Hamiltonian calculated from the full Q-box folded-diagram renormalization \cite{SUN2017227}.
The equation of motion method used along with Gamow states has also been introduced via the in-medium similarity renormalization group method \cite{PhysRevC.99.061302}.

As a consequence, GSM is well suited to study the proton-rich isotonic systems of $^{16}$O, which will be the object of this paper.
These isotones are interesting for several reasons.
Firstly, they are the mirror systems of the neutron-rich isotopes of the oxygen chain,
which is being intensively studied experimentally \cite{PhysRevC.96.054322,PhysRevLett.100.152502,PhysRevC.95.041301,PhysRevC.88.034313} and theoretically \cite{Holt2013,PhysRevLett.108.242501,PhysRevC.96.024308}.
Consequently, one can directly consider its isospin-breaking features, as continuum coupling is known to break isospin symmetry,
in particular with the so-called Thomas-Ehrmann shift \cite{PhysRev.81.412,PhysRev.88.1109}.
Moreover, as noticed in Ref.\cite{PhysRevLett.105.032501,PhysRevLett.110.022502},
an important contribution coming from three-body forces, of a few MeV, has been seen to be necessary to properly reproduce the experimentally known binding energies at either proton or neutron drip-line.
Three-body forces can be rather precisely mimicked by a two-body effective Hamiltonian at neutron drip-line, as the latter method only leads to an overbinding of 500 keV for $^{28}$O in Ref.\cite{PhysRevC.96.024308}.
However, an $A$-dependence of the Hamiltonian had been seen to be necessary in Ref.\cite{DeGrancey:2016bez} to reproduce the $^{14}$O(p,p) elastic reaction cross section, while using the same Hamiltonian structure.
It has thus appeared that approximating three-body forces by a simple two-body operator might not be as precise at proton drip-line as at neutron drip-line.
Finally, the treatment of the Coulomb Hamiltonian can be analyzed rigorously with GSM.
Indeed, the method of Ref.\cite{PhysRevC.83.034325} allows to include the Coulomb Hamiltonian almost exactly, on the one hand,
and the correlations induced by the Coulomb interaction in the asymptotic region are present as well due to the inclusion of continuum coupling, on the other hand.
In particular, one can test the precision of the formula usually used for Coulomb energy, based on the isobaric multiplet mass equation (IMME) \cite{IMME,PhysRevC.55.2407,PhysRevC.74.034315}.

This paper is constructed as follows. We will firstly recall the basic features of GSM, by insisting on the issues raised by the presence of the Coulomb interaction in the asymptotic region.
We will then describe the Hamiltonians used to describe the studied isotones of $^{16}$O, where a core + valence protons approach will be considered.
We will compare two different interactions, the first one being local, of the Furutani-Horiuchi-Tamagaki (FHT) type \cite{FHT1,FHT2}, and the second one non-local,
as issued from effective field theory (EFT) \cite{MACHLEIDT20111}.
Then, we will show the calculated energy spectra of the proton-rich isotones of $^{16}$O, the calculated Coulomb contributions in their ground states and in the excited states of $^{18}$Ne and $^{20}$Mg,
using both GSM and IMME frameworks, and the calculation of the elastic scattering cross sections of $^{18}$Ne(p,p).
The latter calculation will allow, in particular, to state the predictive power of our formalism, as the considered cross sections do not enter the fitting procedure utilized to build the used Hamiltonians.
The conclusion of this work will be made afterwards.

\section {Theoretical model}
\begin{figure}[!htb]
\includegraphics[width=1\columnwidth]{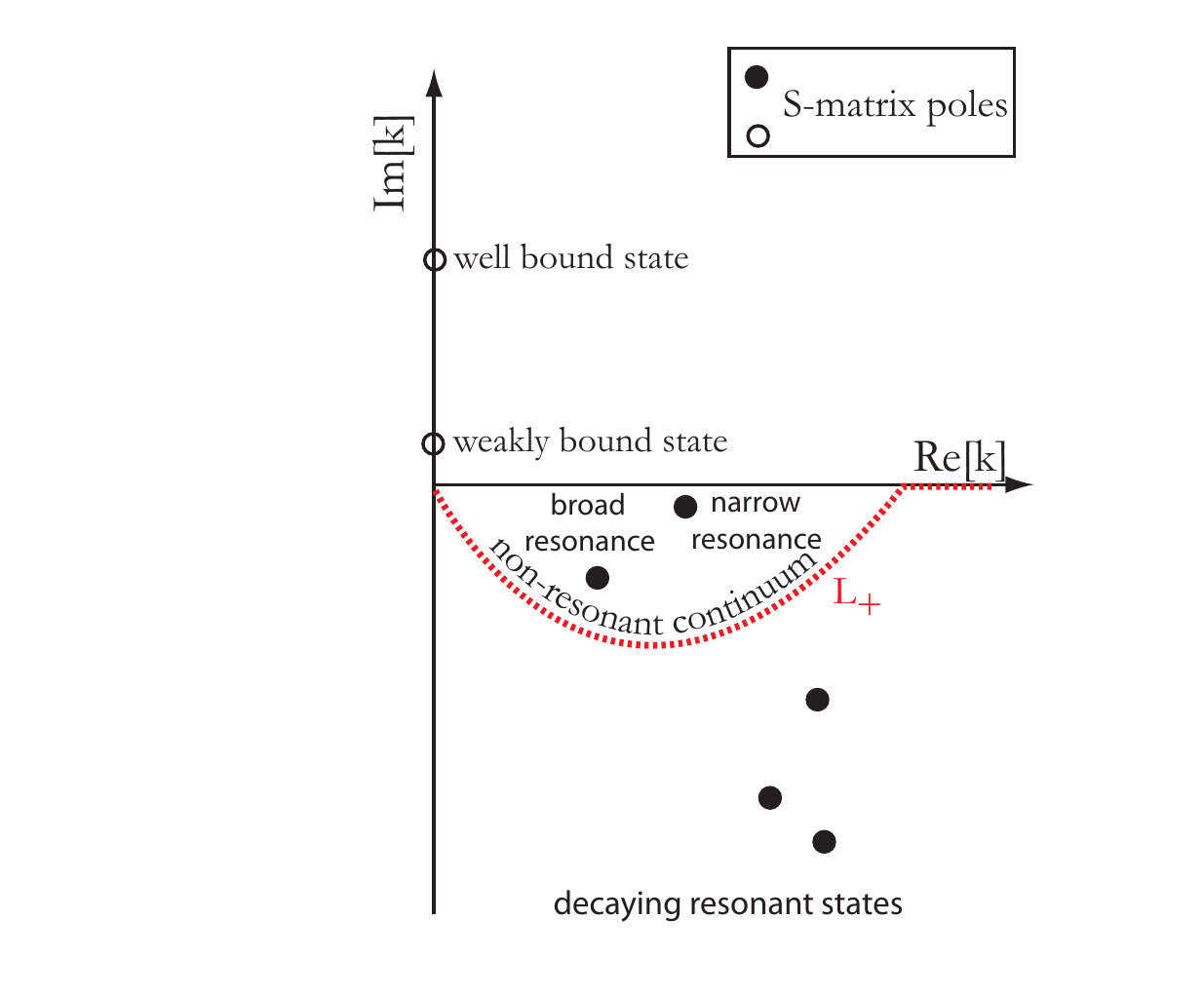}
\caption{
  Depiction of the Berggren basis in the complex momentum plane for a fixed partial wave.
  Typical complex momenta of bound, narrow and broad resonance states, i.e.~S-matrix poles, are provided.
  The $L_+$ contour of scattering states encompasses the S-matrix poles of interest.
  Figure taken from Ref.\cite{PhysRevC.88.044318} (color online).}
\label{Berggren_figure}
\end{figure}

The theoretical model used is that of GSM (see Ref.\cite{0954-3899-36-1-013101}
for a review and Refs.\cite{PhysRevC.94.054302,PhysRevC.96.054316,PhysRevC.96.054322,PhysRevC.96.024308,PhysRevC.91.034609,JPG_GX_Dong,PhysRevC.99.044606} for recent applications of the model).
It is a configuration interaction framework based on the one-body Berggren basis \cite{BERGGREN1968265}.
The Berggren basis is generated by a finite-range potential and possesses bound, resonance and scattering states:
\begin{equation}
\sum_n \ket{n} \bra{n} + \int_{L^+} \ket{k} \bra{k}~dk = 1 \label{Berggren}
\end{equation}
where $\ket{n}$ is a resonant state, and $L^+$ is a contour in the complex momentum plane, starting from $k=0$,
going to $k \rightarrow +\infty$ and encompassing the resonance states of the finite sum (see Fig.(\ref{Berggren_figure})).
Consequently, the Berggren basis is the complex extension of the real-energy completeness relation of Newton \cite{newton1966scattering}, consisting of bound states and of a continuum of real-energy scattering states.
Contrary to the Newton completeness relation, with which only localized states can be expanded, the Berggren basis can expand resonance states.
Equation(\ref{Berggren}) is one-dimensional, i.e.~it is built from the one-body states of a fixed partial wave.
Thus, the three-dimensional completeness relation at one-body level is recovered by considering Eq.(\ref{Berggren}) for all possible partial waves.
The many-body completeness relation follows by constructing Slater determinants from the bound, resonance and scattering Berggren one-body basis states \cite{0954-3899-36-1-013101}.
From a numerical point of view, the $L^+$ contour is discretized with a Gauss-Legendre quadrature in order to obtain an eigenproblem similar to that of HO-SM \cite{0954-3899-36-1-013101}.
It has been checked that 15-45 states per contour are necessary to have converged results.
The Hamiltonian is represented by a complex symmetric matrix, which must be diagonalized.
This is handled efficiently with the complex symmetric extension of the Jacobi-Davidson method \cite{Jacobi_Davidson},
where one takes advantage of the relatively small coupling to the continuum to have a fast convergence of eigenstates.
While the full configuration space is extremely large, reasonably small dimensions are obtained by truncating the GSM space with two scattering states occupied at most in the continuum, which we denote as 2p-2h.
It is possible to lift space truncation if one uses the density renormalization group method \cite{PhysRevLett.97.110603,PhysRevC.79.014304},
as the many-body basis of Slater determinants is hereby replaced by a correlated many-body basis, issued from the diagonalization of the density matrix.
However, continuum coupling is usually small in the $A \sim 20$ region \cite{PhysRevC.67.054311}, so that 2p-2h truncations are sufficiently precise.

The presence of the Coulomb Hamiltonian generates additional difficulties when using a Berggren basis of proton states.
Firstly, the asymptote of one-body basis states is that of Coulomb wave functions, whose structure is more complex than that of the Hankel functions occurring in the neutron case.
The use of the computational methods of Ref.\cite{MICHEL2007232} are hereby necessary to precisely calculate bound, resonance and scattering states with a Coulomb wave function asymptote.
Secondly, the Hamiltonian is of infinite range due to the Coulomb Hamiltonian. In order to deal with this situation, we use the method developed in Ref.\cite{PhysRevC.82.044315}.
For this, one separates the Coulomb Hamiltonian into a one-body potential part, whose asymptote is in $(Z-1)/r$, to which a finite-range residual two-body part is added.
The residual two-body Coulomb part is expanded with a basis of HO states, so that it can be treated in a standard fashion using the Talmi-Brody-Moshinsky transformation \cite{Talmi,MOSHINSKY1959104}.
The one-body potential part of the Coulomb Hamiltonian can be directly inserted in the basis-generating potential. Its treatment is then exact as done at basis level.
However, one cannot use this method if we consider observables involving nuclear eigenstates of different charges, as will be the case for the $^{18}$Ne(p,p) cross sections.
Indeed, in this case, the one-body potential part generates infinite matrix elements using a Berggren basis.
The solution to this problem, detailed in Ref.\cite{PhysRevC.83.034325}, consists in replacing the latter infinite matrix elements by large but finite matrix elements,
depending on the used discretization for the $L^+$ contour (see Fig.(\ref{Berggren_figure})).
Consequently, GSM can be used without problems with proton Berggren basis states.

The used GSM approach is that of a core + valence protons picture. The core used is that of an $^{16}$O core, which is mimicked by a Woods-Saxon (WS) potential. 
All partial waves up to $\ell = 3$ are taken into account.
As the centrifugal barrier increases quickly with orbital angular momentum,
the effect of the associated partial waves on wave function asymptotes is negligible.
Hence, we use the Berggren basis for $spd$ partial waves, whereas $f$ partial waves are represented by a basis of harmonic oscillator states.
Two different residual nuclear proton-proton interactions have been considered.
The first one is that of the FHT interaction, which is Gaussian-based and bears central, spin-orbit and tensor parts.
Their coupling constants, written as $V_c^{S}$,$V_{\scriptscriptstyle{L\:\!\!S}}^{S}$ and $V_{T}^{S}$, respectively, are function of the spin of the two protons, equal to $S=0,1$.
The FHT interaction has been used in GSM to describe the structure of light neutron-rich nuclei in the $A=4-10$ region \cite{PhysRevC.96.054316}, to study neutron-rich oxygen isotopes \cite{PhysRevC.96.024308},
and to evaluate radiative capture reactions in $A=6-8$ nuclei \cite{PhysRevC.91.034609,JPG_GX_Dong}.
The second interaction is based on the EFT formalism \cite{MACHLEIDT20111}.
It arises from an order by order expansion of contact terms in the EFT interaction,
whose parameters are fitted on the experimental data associated to the structure of proton-rich nuclei \cite{PhysRevC.98.044301}.
The EFT interaction possesses the overall features of realistic nucleon-nucleon interactions \cite{SUN2017227}, and is non-local in the short-range region, contrary to the FHT interaction.
We will see, in fact, that the results of the EFT interaction are closer to experimental data than those provided by the FHT interaction.

A dependence on the number of nucleons has been added to the WS potential of the core and to the FHT and EFT two-body interactions.
Indeed, it mimicks the effect of missing three-body forces \cite{PhysRevLett.90.042502,PhysRevC.59.R2347}.
We will see in the next section that an $A$-dependence of the Hamiltonian is necessary to reproduce experimental energies.
The $A$-dependence used for the Hamiltonian is standard \cite{Bohr_Mottelson,shell_model_review}:
\begin{eqnarray}
F_{1b} &=& 1 + f \left( \frac{N - Z}{A} \right) \label{one_body_A_dep} \\
F_{2b} &=& \left( \frac{A_{core} + 2}{A} \right)^e \label{two_body_A_dep}
\end{eqnarray}
where $Z$,$N$ and $A$ are the number of protons, neutrons and nucleons of the nucleus, $A_{core}$ is the number of nucleons of the core, here equal to 16,
$F_{1b}$ is the $A$-dependent one-body factor multiplied to the central depth of the WS potential, depending on the parameter $f$,
and $F_{2b}$ is the $A$-dependent two-body factor multiplied to the matrix elements of the used effective interaction, depending on the exponent parameter $e$.

%%%%
\begin{table}[htb] 
\centering
\caption{\label{Table.InterParamFHT} The optimized parameters of the FHT interaction consist of central ($V_c^{S}$),
  spin-orbit ($V_{\scriptscriptstyle{L\:\!\!S}}^{S}$) and tensor ($V_T^{S}$) coupling constants \cite{PhysRevC.96.054316}. $S=0,1$ is the spin of the two protons.}
\begin{tabular}{l|ccccc}\hline \hline 
Parameter &$V_c^{1}$  & $V_c^{0}$    &$V_{\scriptscriptstyle{L\:\!\!S}}^{1}$   & $V_{T}^{1}$  \\ \hline 
Value&  $-$2.00 &  $-$7.73  &  17.86 &  $-$56.79 \\ \hline \hline
\end{tabular}
\end{table}

The Hamiltonian parameters have been fitted on the single-particle states of $^{17}$F, and on the known experimental energies of the low-lying states of the $Z=10-12$ isotones of $^{16}$O.
The $f$ and $e$ parameters of Eqs.(\ref{one_body_A_dep},\ref{two_body_A_dep}) have been fixed to 0.043 and 0.3, respectively, when using $A$-dependent interactions.
The parameters of the WS core potential are the same using either the FHT or EFT interactions.
They consist of the diffuseness $d = 0.65$ fm, radius $R_0 = 2.98$ fm,
central strength $V_0 = 56$ for $\ell = 0$, $V_0 = 58.075$ for $\ell = 1,3$ and $V_0 = 58.003$ for $\ell = 2$,
and spin-orbit strength $V_{ls} = 6.539$ for $\ell = 1,3$ and $6.5$ for $\ell = 2$.
The parameters of the FHT and EFT interaction are shown in Tabs.(\ref{Table.InterParamFHT},\ref{Table.InterParamEFT}), respectively.
The EFT parameters consist of the leading order parameters, denoted as $C_S$ for its spin-independent part and $C_T$ for its spin-dependent part,
and of the next-to-leading order parameters, denoted as $C_{1 \dots 7}$. The latter notation for constants is standard (see Ref.\cite{MACHLEIDT20111} for details).
The EFT interaction is a low-momentum expansion valid for momenta much smaller than 1 GeV \cite{MACHLEIDT20111}.
Therefore, following the prescription of Ref.\cite{MACHLEIDT20111}, the EFT interaction is renormalized by way of a momentum-dependent regulator function $f(p,p')$:
\begin{eqnarray}
V_{EFT}(\mathbf{p'},\mathbf{p}) &\rightarrow& V_{EFT}(\mathbf{p'},\mathbf{p})f(p',p) \label{EFT_renormalization} \\
f(p',p) &=& \exp \left( -\left( \frac{p'}{\Lambda} \right)^{2n} - \left( \frac{p}{\Lambda} \right)^{2n} \right) \label{EFT_regulator}
\end{eqnarray}
where $\Lambda$ is a cut-off energy of 300 MeV, and $n=2,3,4$ according to the LO or NLO constant and two-nucleon partial wave considered (see Ref.\cite{MACHLEIDT20111} for details).

The fitted EFT parameters lie between 0.13 and 2.46 in absolute value, so that they follow the naturalness condition of low energy constants, i.e.~that they are expected to be close to unity \cite{MACHLEIDT20111}.
%%%%
\begin{table}[htb]
\centering
\caption{\label{Table.InterParamEFT} Optimized parameters of the EFT interaction at leading order (LO) and next-to-leading order (NLO), in natural units. See Ref.\cite{MACHLEIDT20111} for definitions and notations.}
\begin{tabular}{l|cccccccc} \hline \hline
LO constant & $C_S$    & $C_T$   \\  \hline 
LO value&  $-$2.46  & $-$0.60 \\ \hline  \hline 
NLO constant &$C_1$   & $C_2$    & $C_3$  & $C_4$  & $C_5$   & $C_6$  & $C_7$ \\ \hline 
NLO value&  $-$0.25  &  0.55  &  $-$0.25 & $-$0.29 & $-$0.46 & $-$0.13 & $-$0.49  \\ \hline \hline
\end{tabular}
\end{table}

\section {Spectra}
\begin{figure}[!htb]
\includegraphics[width=1\columnwidth]{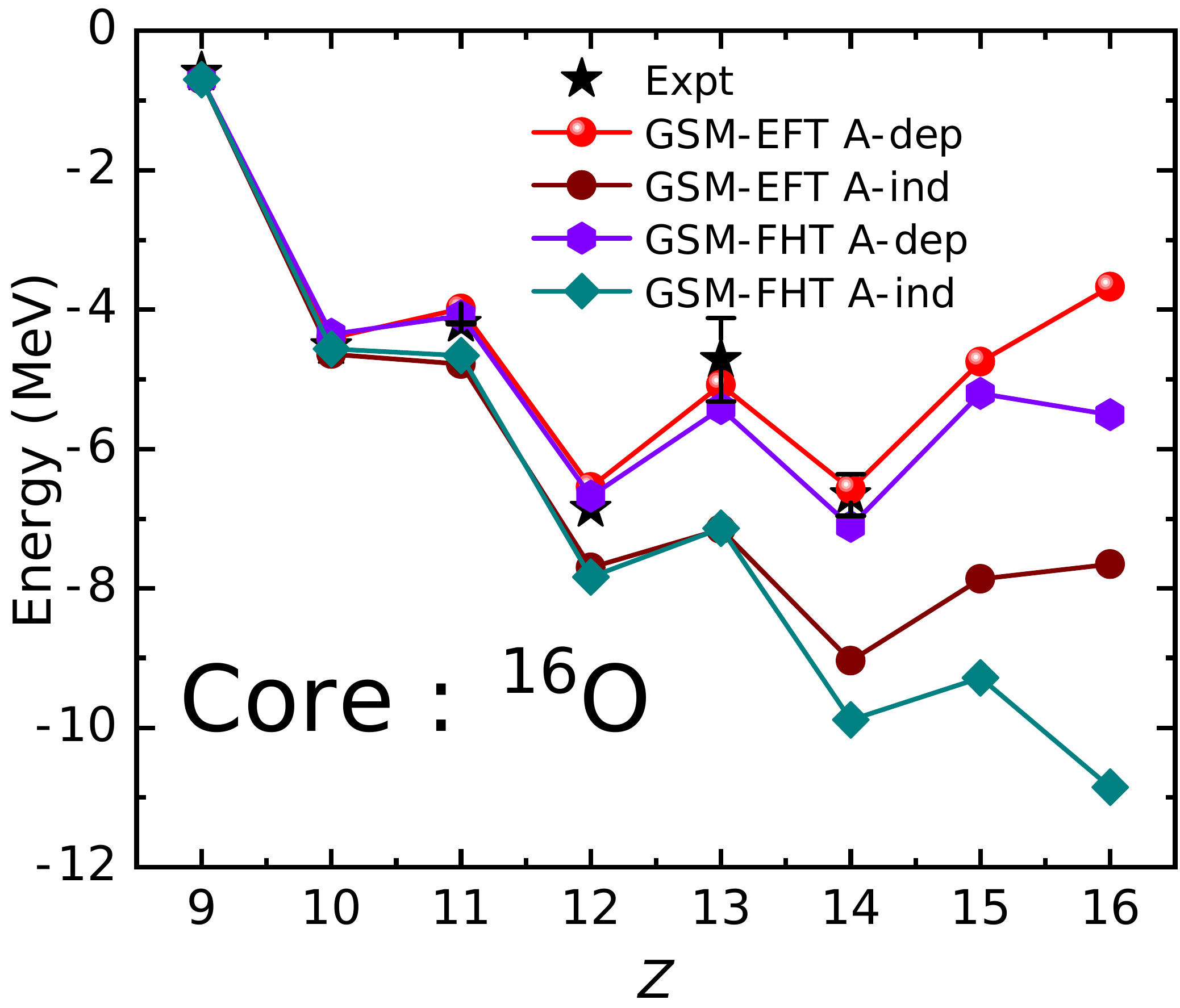}
\caption{Theoretical and experimental binding energies (in MeV) of the $^{16}$O isotones with respect to the $^{16}$O core as a function of their number of protons $Z$.
  Theoretical binding energies arise from a GSM calculation using the EFT or FHT interaction, using $A$-dependent interactions (A-dep) or $A$-independent interactions (A-ind).
  Experimental data are denoted by stars and are taken from AME2016 \cite{wang2017ame2016} (color online).
}
\label{Binding_E}
\end{figure}

The theoretical and experimental binding energies of the isotones of $^{16}$O from $^{17}$F to $^{22}$Si are shown in Fig.(\ref{Binding_E}).
One can see that the Hamiltonians used separate into two parts, with one consisting of $A$-independent core potential and valence nucleon-nucleon interaction,
and the other comprising $A$-dependence in the parameters defining the one-body and two-body parts.
While $A$-dependent Hamiltonians provide with a good description of experimental binding energies, $A$-independent Hamiltonians exhibit a very strong overbinding of energies when $A$ increases.
The discrepancy is already of a few hundred of keV for $^{19}$Na, and worsens along with the number of valence protons, to reach around 2.5 MeV for $^{22}$Si.
It is also present using both EFT and FHT interactions, where one can note that the overbinding encountered with the FHT interaction is larger than that of the EFT interaction.
In fact, the only nucleus whose binding energy is well reproduced is $^{18}$Ne.
Such a behavior had already been noticed in Ref.\cite{PhysRevLett.110.022502} for the same nuclear systems.
Moreover, the description of the $^{14}$O(p,p) reaction with GSM \cite{DeGrancey:2016bez}, involving the unbound proton-rich nucleus $^{15}$F,
also demanded $A$-dependence in the Hamiltonian to properly account for proton-emission widths of the unbound spectra of $^{14}$O and $^{15}$F.
Hence, it is highly probable that the need for strong three-body forces among valence nucleons, or, equivalently, $A$-dependent Hamiltonians, is a common feature of nuclei at proton drip-line.

Let us compare the ground states of $^{16}$O isotones to their isobaric analogue states, i.e.~neutron-rich oxygen isotopes, by replacing valence protons by valence neutrons.
In order for the oxygen chain to be well reproduced with realistic interactions, three-body forces must be included for energy saturation to occur at neutron drip-line \cite{Holt2013}.
Nevertheless, three-body forces can be usually  mimicked by an equivalent phenomenological two-body interaction at neutron drip-line.
Indeed, using a WS potential + $A$-independent FHT interaction with GSM in this region has been noticed to be only responsible for an overbinding of the ground state of $^{28}$O close to 500 keV \cite{PhysRevC.96.024308}. 
Added to that, the isobaric analogue states of the $^{16}$O isotones bearing $A=17-22$ nucleons,
consisting of $^{17-22}$O, can be very well reproduced with a phenomenological two-body interaction, in HO-SM \cite{PhysRevLett.105.032501} or in GSM \cite{PhysRevC.67.054311}.

One observes that the saturation occurring at the neutron drip-line of oxygen isotopes does not exist in corresponding isotones.
A strong odd-even staggering continues instead to develop therein even well after the proton drip-line, attained with $^{19}$Na.
More precisely, the experimental binding energies of oxygen isotopes differ by about 500 keV at neutron drip-line, whereas they can differ by 2-3 MeV in proton-rich isotones \cite{ensdf}.
The occurring strong odd-even staggering could be explained if protons are sufficiently localized in the nuclear zone, as they could generate sizable pairing energy in this situation.
\begin{figure}[!htb]
\includegraphics[width=1\columnwidth]{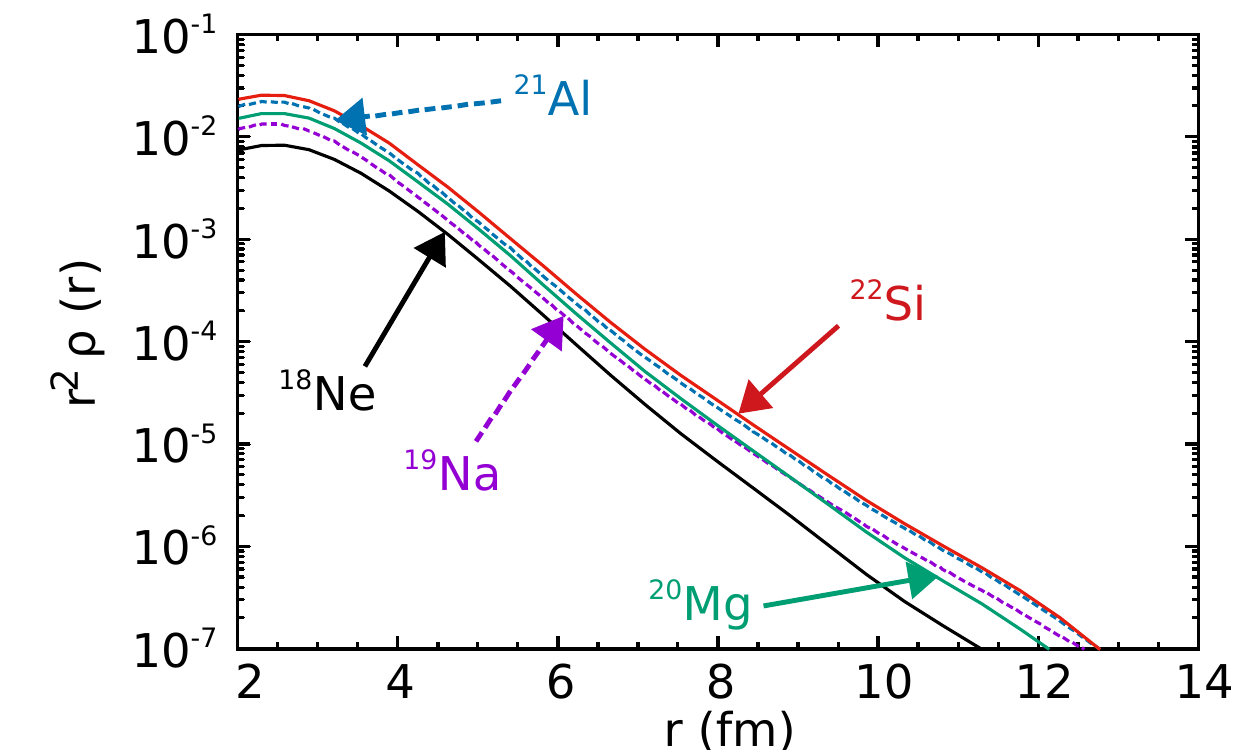}
\caption{Radial densities multiplied by radius squared, denoted as $r^2 \rho(r)$, of the ground states of the $^{16}$O isotones for $A=18-22$ as a function of the radius $r$ (in fm).
Even and odd isotones are represented with solid and dashed lines, respectively, and are indicated by arrows (color online).}
\label{radial_densities_ground_states}
\end{figure}
This assumption can be considered by calculating the proton radial densities of the studied ground states.
One can note that densities are complex for resonance states, as these eigenstates have a complex energy.
However, their imaginary part is small compared to their real part in our calculations, so that we will only consider real parts when dealing with densities in this paper.
It happens that proton radial densities decay rapidly for all isotones, even if they are unbound (see Fig.(\ref{radial_densities_ground_states})).
In fact, the density of resonance states will start increasing at a radius larger than 14 fm (see Fig.(\ref{radial_densities_ground_states})).
Therefore, one can consider that all many-body wave functions behave as bound states for $r < 14$ fm.
When using a basis of harmonic oscillator states, as in Ref.\cite{PhysRevLett.110.022502},
this phenomenon might be seen as arising only from the well-bound character of the basis, artificially localizing many-body wave functions.
As continuum coupling is exactly taken into account in GSM, the asymptotes of many-proton systems are precisely evaluated, so that this artifact is nonexistent in GSM.
Consequently, the enhanced proton-proton correlations at proton drip-line are very probably due to the confining effect of the Coulomb barrier. 
The important three-body effect would then be explained by the presence of the Coulomb interaction.
Let us consider only the two-body part of the Hamiltonian to state this fact.
As the two-body part of the nucleon-nucleon interaction is mostly attractive and the Coulomb interaction repulsive,
the total matrix elements of the two-body part of the Hamiltonian are generally smaller in absolute value for protons than for neutrons.
Consequently, the effect of three-body forces is comparatively larger for proton-rich nuclei than for neutron-rich nuclei.
The replacement of three-body forces by a renormalized two-body interaction of simple form then appears to be too strong an approximation at proton drip-line.
Two methods have successfully tackled the three-body part of the nuclear Hamiltonian in the proton-rich region.
The first method, used in Ref.\cite{PhysRevLett.110.022502} within many-body perturbation theory (MBPT), consists to calculate directly the one-body and two-body parts of the three-body force using normal ordering.
As the residual part of three-body forces is negligible \cite{PhysRevC.76.034302}, this method replaces the initial three-body force by a very close two-body operator,
providing in particular with the induced monopole correction \cite{PhysRevC.59.R2347}.
The second method, utilized in Ref.\cite{DeGrancey:2016bez} and this paper, consists in adding an $A$-dependence to the used one-body and two-body parts of the Hamiltonian.
It is indeed equivalent to monopole correction \cite{PhysRevLett.90.042502}.

\begin{figure}[!htb]
\includegraphics[width=1\columnwidth]{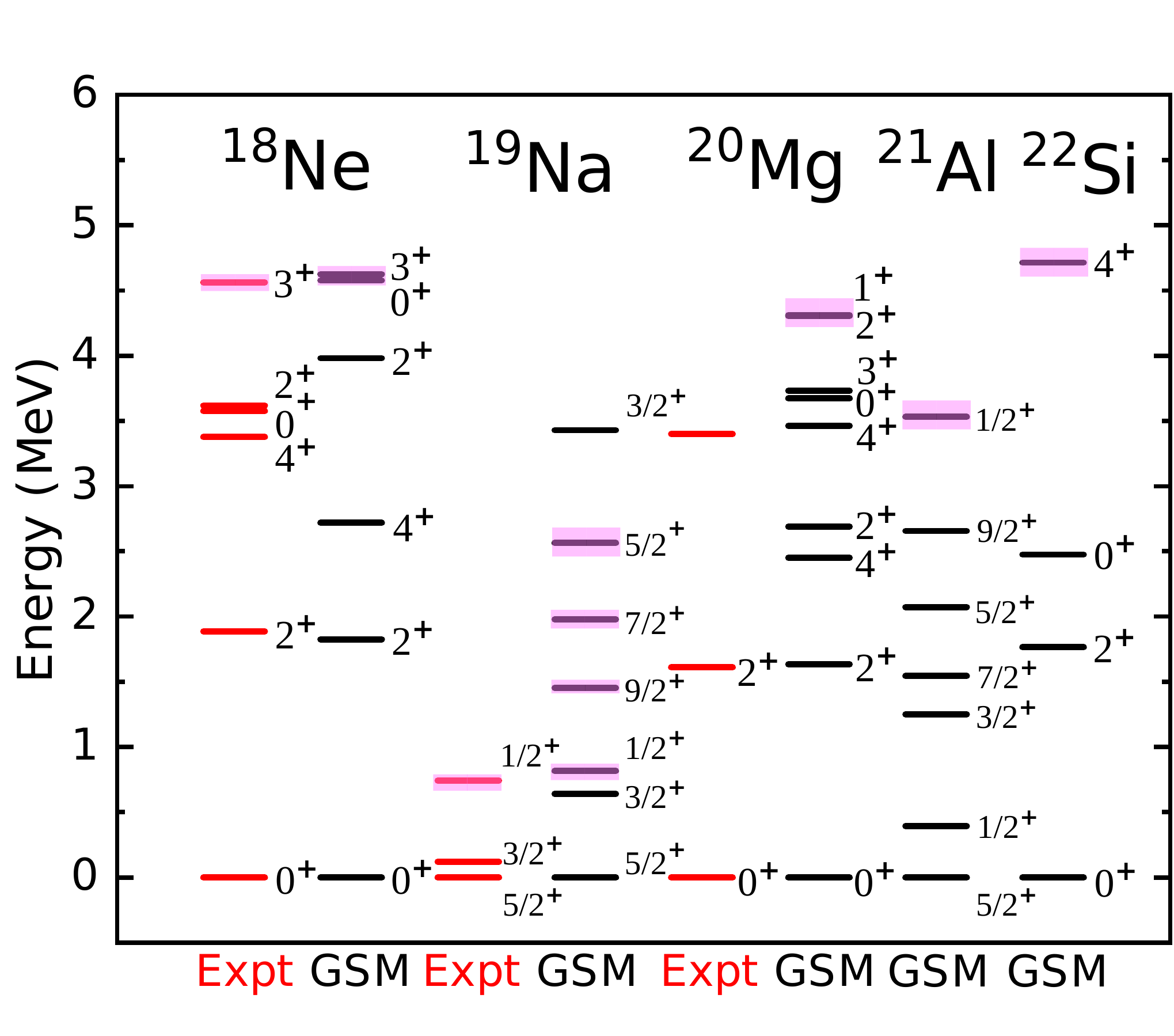}
\caption{Theoretical and experimental spectra of $^{16}$O isotones for $A=18-22$. Energies are given in MeV. Proton-emission widths, also given in MeV, are represented by shaded areas.
Theoretical energies and widths have been calculated with GSM using the $A$-dependent EFT interaction. Experimental data come from Ref.\cite{PhysRevLett.110.022502,PhysRevC.67.014308,PhysRevC.82.054315,PhysRevC.76.024317,ensdf}.}
\label{spectra}
\end{figure}

The spectra of proton-rich nuclei from $^{18}$Ne to $^{22}$Si, obtained using the EFT interaction in GSM, are depicted in Fig.(\ref{spectra}) along with available experimental data.
The spectrum of $^{18}$Ne is qualitatively well described, considering that it bears only two valence protons. Indeed, the energy of the $2^+_1$ state is very well reproduced.
Its $4^+_1$, $2^+_2$ and $3^+_1$ excited states differ from experimental data by a few hundreds of keV at most, while the calculated $0^+_2$ excited state is too high in energy by about 1 MeV.
The spectrum of $^{19}$Na is of the same quality, as the energy of the $1/2^+_1$ state
is very close to experimental data, while that of the $3/2^+_1$ state is too low by about 500 keV.
The spectra of the same nuclei obtained with MBPT in Ref.\cite{PhysRevLett.110.022502} are comparable. Indeed, while the energy of the $4^+_1$ state of $^{18}$Ne is closer to experiment therein,
that of the $2^+_2$ state of $^{18}$Ne is about 1 MeV too low.
The GSM calculation of the first excited state of $^{20}$Mg, of $2^+$ character, compares well with experiment.
The second excited state of $^{20}$Mg, predicted to be a $4^+$ state in Ref.\cite{PhysRevC.99.021301}, would be too low with GSM, similarly to that of $^{18}$Ne.
The spectra of $^{21}$Al and $^{22}$Si, not known experimentally, can be compared to those obtained in Ref.\cite{PhysRevLett.110.022502}.
Our first excited states are in the same order, except for the $7/2^+_1$ state of $^{21}$Al, slightly lower in our case.
Hence, the comparison of our GSM results with experimental data, on the one hand, and the theoretical calculations of Ref.\cite{PhysRevLett.110.022502}, done with the realistic framework of MBPT, on the other hand,
showed that the GSM Hamiltonian can provide with physically sound excited states.
The spectra provided by the FHT interaction do not compare well with experimental data, however, so that they are not shown.  
Consequently, one can properly estimate the Coulomb contribution of the Hamiltonian to the excited states using the $A$-dependent EFT interaction and compare it to that effected in other theoretical models.
We will also only consider the $A$-dependent EFT interaction with GSM afterwards.

\section {Coulomb contributions}
\begin{figure}[!htb]
\includegraphics[width=1\columnwidth]{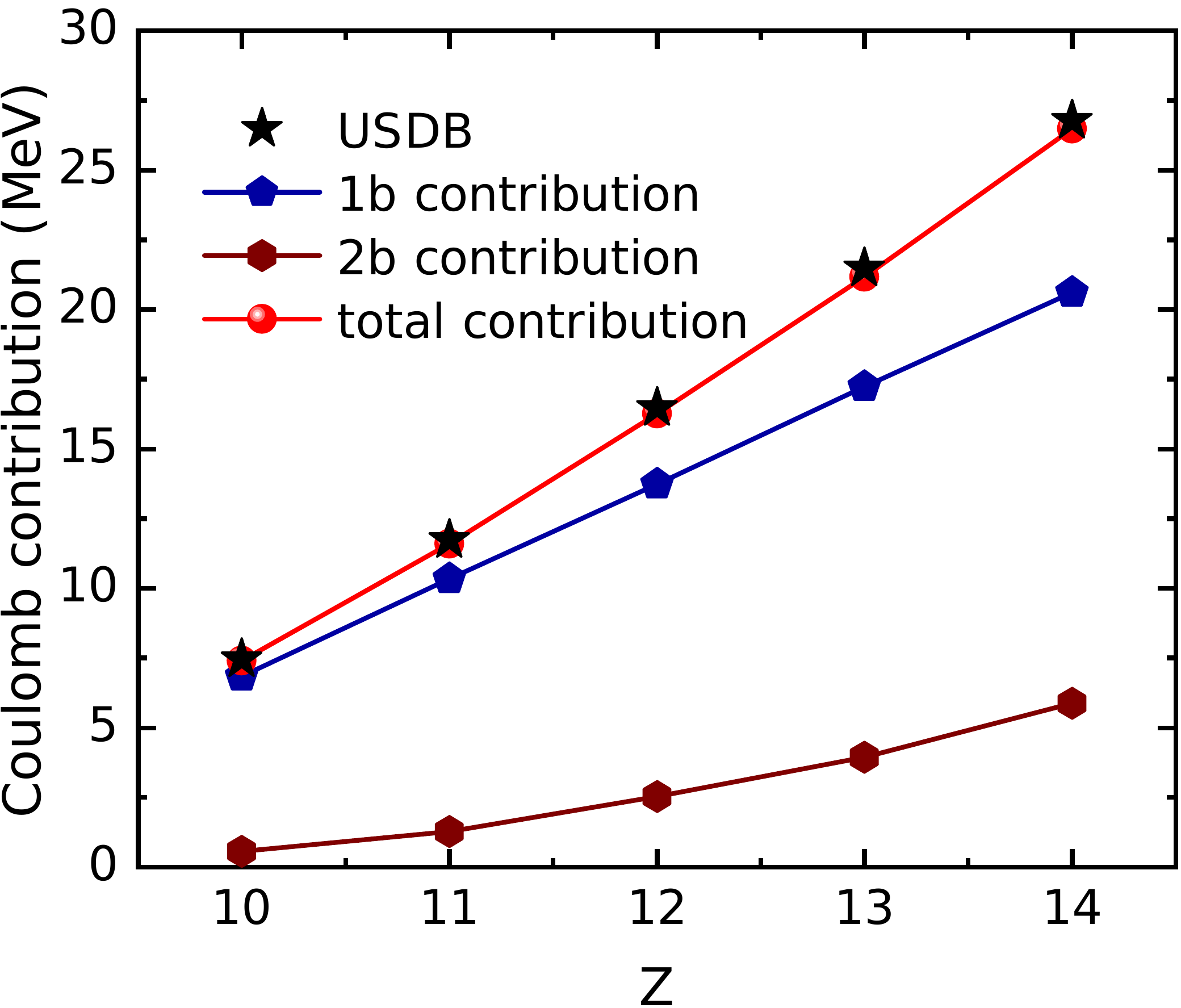}
\caption{Hamiltonian Coulomb contributions (in MeV) of the ground states of $^{16}$O isotones as a function of their proton number $Z$.
  Solid lines are issued from a GSM calculation using the $A$-dependent EFT interaction. Stars point out the IMME Coulomb energies \cite{IMME,PhysRevC.55.2407,PhysRevC.74.034315}.
  Shown GSM results consist of the $^{16}$O core Coulomb one-body part (1b contribution), of the valence Coulomb two-body part (2b contribution), and of their sum (total contribution).
  Solid lines are present only to guide the eye (color online).}
\label{coulomb_1}
\end{figure}

\begin{figure}[!htb]
\includegraphics[width=1\columnwidth]{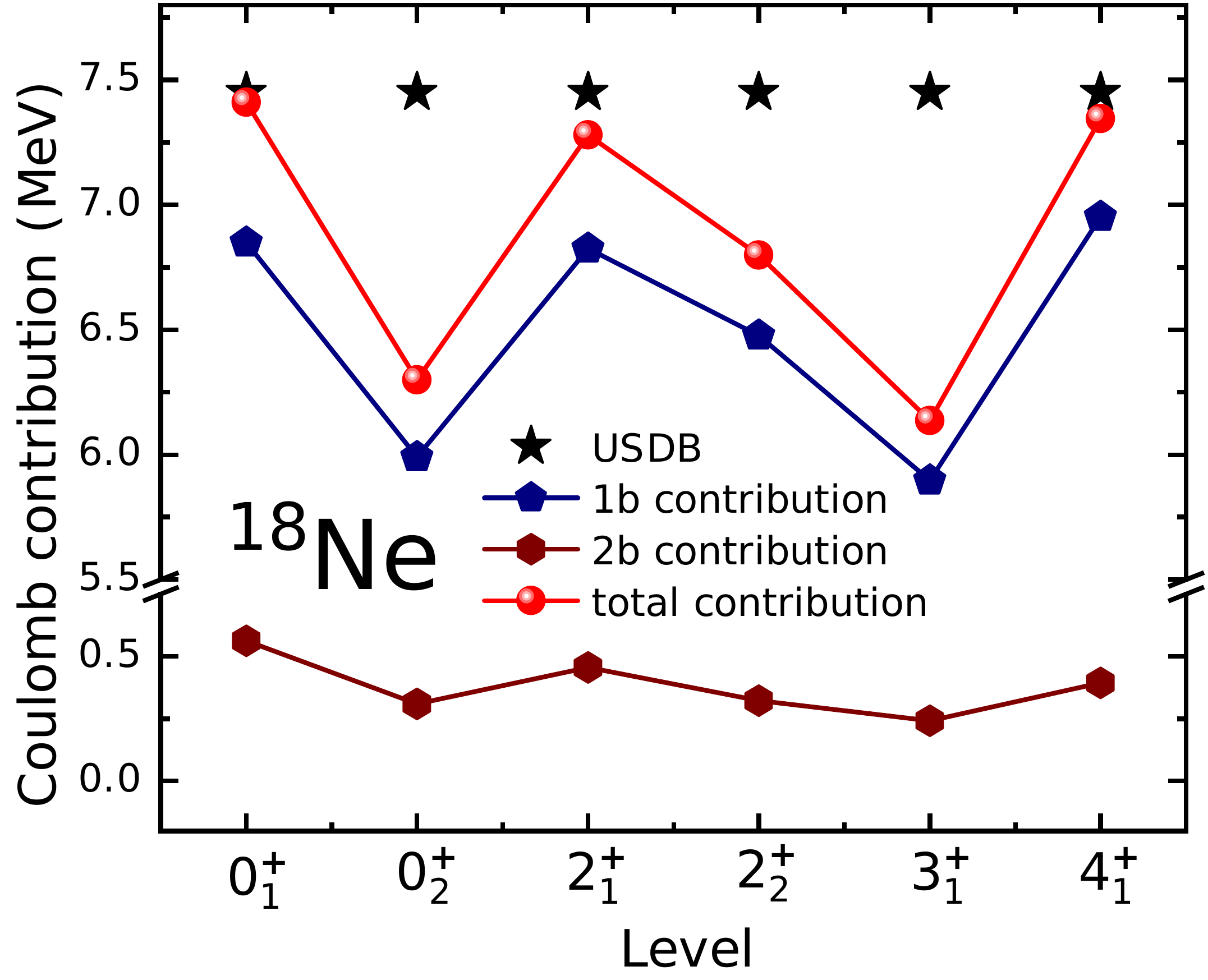}
\caption{Same as Fig.(\ref{coulomb_1}), but for the ground state and excited states of $^{18}$Ne calculated with GSM using the $A$-dependent EFT interaction (see Fig.(\ref{spectra})) (color online).}
\label{coulomb_2}
\end{figure}

\begin{figure}[!htb]
\includegraphics[width=1\columnwidth]{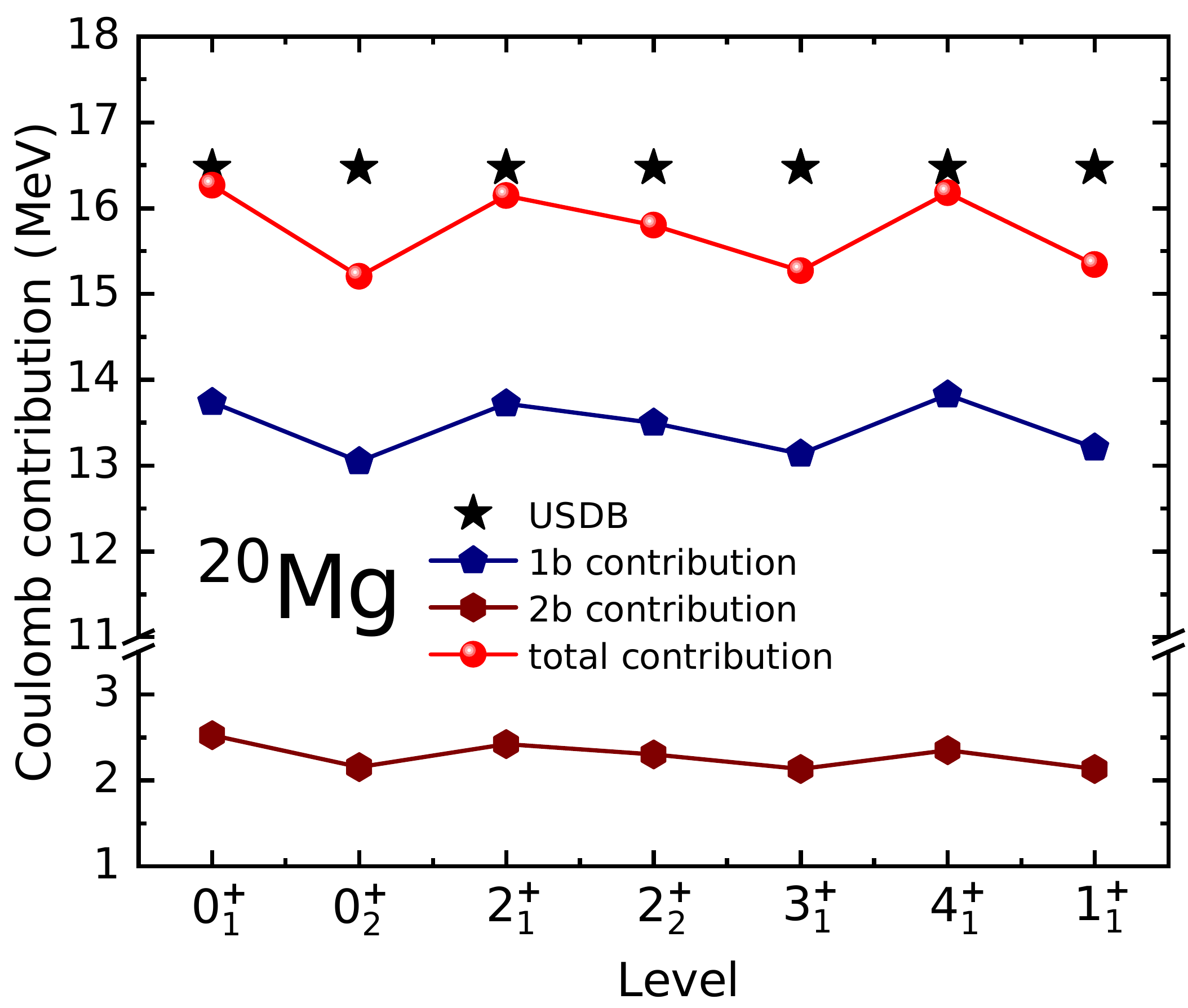}
\caption{Same as Fig.(\ref{coulomb_2}), but for the ground state and first excited states of $^{20}$Mg (see Fig.(\ref{spectra})) (color online).}
\label{coulomb_3}
\end{figure}

The Coulomb contributions, i.e.~the expectations values of the one-body, two-body and total Coulomb Hamiltonians are depicted in Fig.(\ref{coulomb_1})
for the ground states of $^{16}$O isotones from $^{18}$Ne to $^{22}$Si, and in Figs.(\ref{coulomb_2},\ref{coulomb_3}) for the spectra of $^{18}$Ne and $^{20}$Mg, respectively.
The IMME formula used along with the USDB interaction should depend in principle on angular momentum and state number \cite{PhysRevC.55.2407}.
However, in practice, the IMME values are function of isospin only \cite{PhysRevC.55.2407},
so that they are equal for all the Hamiltonian eigenstates of a fixed nucleus when using only proton valence states (see Figs.(\ref{coulomb_2},\ref{coulomb_3})).
%%%%
%%%%
%%%%%%%
On Fig.(\ref{coulomb_1}), one can see that the Coulomb contributions follow the usual rules present in closed quantum systems.
Indeed, its one-body part grows linearly with the number of protons, while its two-body part has a quadratic increase.
Moreover, the total Coulomb contribution is almost equal to the IMME value.
This phenomenon is consistent with the observation done in the previous section when considering odd-even staggering and densities,
i.e.~that the Coulomb Hamiltonian has a tendency to confine valence protons inside the nuclear zone even if wave functions are unbound.
The IMME formula is no longer suitable for excited states, however (see Figs.(\ref{coulomb_2},\ref{coulomb_3})).
Indeed, the total Coulomb contribution differs by several hundreds of keV to more than 1 MeV when going from the ground states to neighboring excited states.
In $^{18}$Ne (see Fig.(\ref{coulomb_2})), while the first three states of the spectrum have a comparable Coulomb energy, it is smaller by 0.7 to 1.5 MeV in higher excited states.
As all states lying above the $4^+_1$ state in $^{18}$Ne are unbound, one can infer that the drop in Coulomb energy might be due to the larger extension of these states.
%%%%
\begin{figure}[!htb]
\includegraphics[width=1\columnwidth]{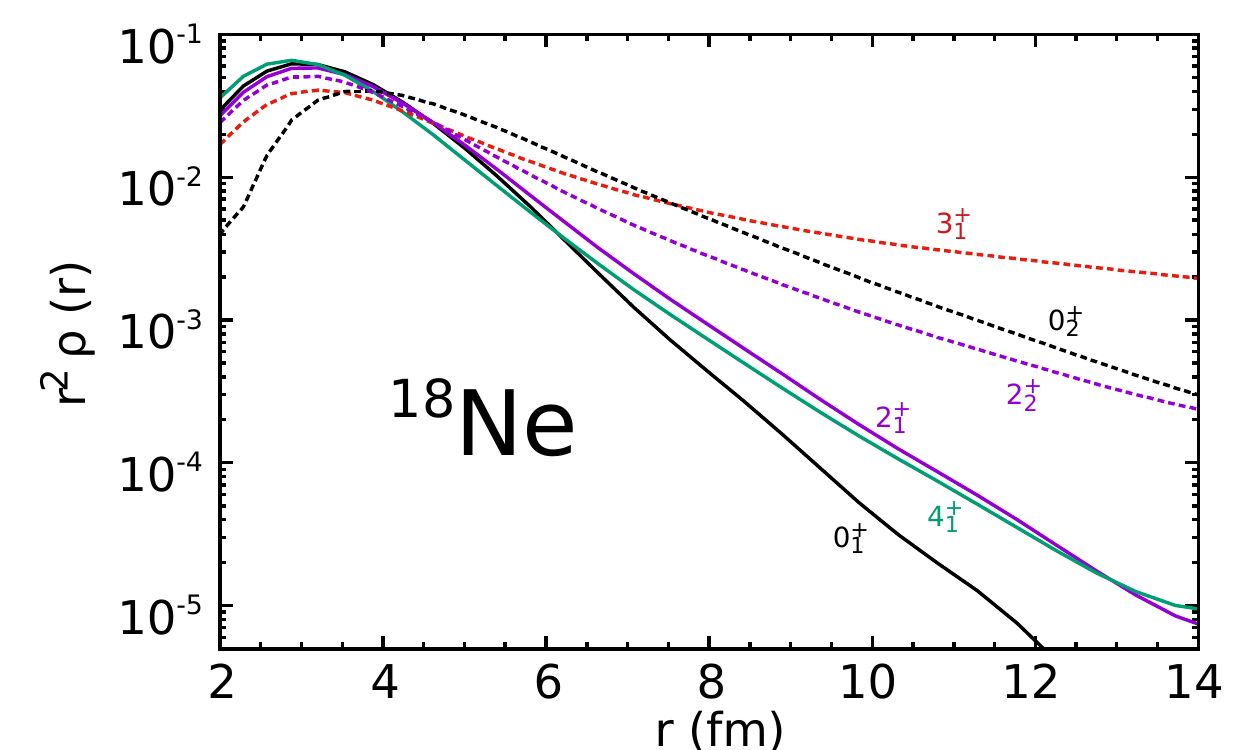}
\caption{Same as Fig.(\ref{radial_densities_ground_states}), but for the ground state and excited states of $^{18}$Ne calculated with GSM using the $A$-dependent EFT interaction (see Fig.(\ref{spectra})).
The bound and unbound states of the spectrum of $^{18}$Ne are represented with solid and dashed lines, respectively (color online).}
\label{radial_densities_18Ne}
\end{figure}
%%%%
\begin{figure}[!htb]
\includegraphics[width=1\columnwidth]{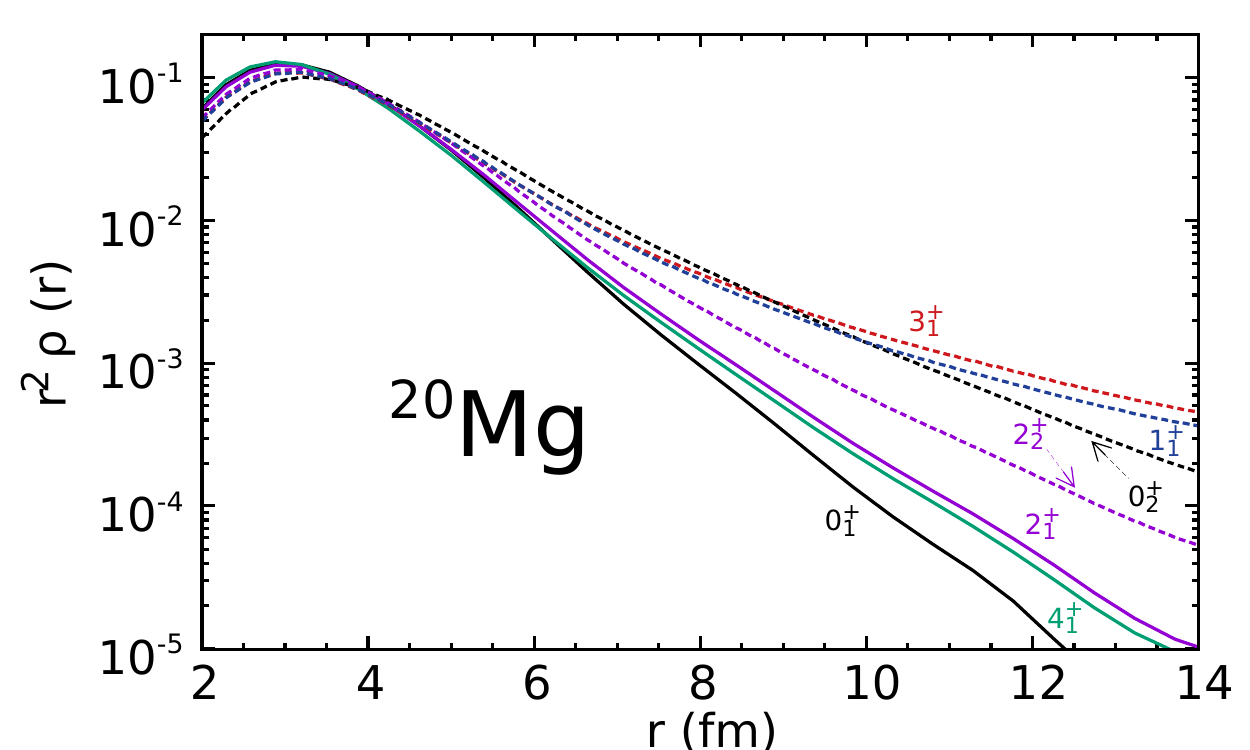}
\caption{Same as Fig.(\ref{radial_densities_18Ne}), but for the ground state and first excited states of $^{20}$Mg calculated with GSM using the $A$-dependent EFT interaction (see Fig.(\ref{spectra})) (color online).}
\label{radial_densities_20Mg}
\end{figure}
%%%%
This fact is corroborated by the asymptotic behavior of the radial densities of the different states of the spectrum of $^{18}$Ne and $^{20}$Mg (see Figs.(\ref{radial_densities_18Ne},\ref{radial_densities_20Mg})).
Indeed, the Coulomb contribution is the smallest for the most extended states, which are the highest in the spectrum as well (see Fig.(\ref{spectra})).
One can note that the $2^+_2$ states of $^{18}$Ne and $^{20}$Mg are intermediate between well bound and well unbound states (see Fig.(\ref{radial_densities_20Mg})) .
This translates into a $2^+_2$ Coulomb contribution which is half way between those of the well bound and unbound states of considered nuclei (see Figs.(\ref{coulomb_2},\ref{coulomb_3})).
Continuum coupling is thus necessary to properly account for the slow decay of unbound states in the medium asymptotic region.
Moreover, one can see that the one-body part of the Coulomb Hamiltonian is dominant, its two-body part only adding a value around 200 to 500 keV for $^{18}$Ne (see Fig.(\ref{coulomb_2})).
Nevertheless, the pattern followed by the one-body and two-body parts of the Coulomb contribution is the same.
For example, the Coulomb contribution of both parts decreases when going from the $0^+_1$ state to the $0^+_2$ state, and augments from the $0^+_2$ state to the $2^+_1$ state (see Fig.(\ref{coulomb_2})).
Consequently, the very different Coulomb contributions are mainly due to the Coulomb interaction of the valence protons with the core.
The situation is qualitatively similar in $^{20}$Mg for the total Coulomb contributions (see Fig.(\ref{coulomb_3})), as Coulomb energies differ by 0.5 to 1 MeV.
One can note that the one-body and two-body parts of the expectation values of the Coulomb Hamiltonian are typically 2 and 4 times larger than in $^{18}$Ne.
It was to be expected because $^{18}$Ne and $^{20}$Mg have 2 and 4 valence protons, respectively.
The contributions of the Coulomb one-body and two-body parts also follow the same pattern in $^{20}$Mg as that described in $^{18}$Ne (see Figs.(\ref{coulomb_2},\ref{coulomb_3})).
However, the two-body Coulomb expectation values in $^{20}$Mg exhibits larger differences than in $^{18}$Ne, so that the variations of total Coulomb contribution are more pronounced than those of its one-body part.
As consequence, we could show that the IMME formula, devised from isobaric analogue states near the valley of stability, is justified only for ground states.
Indeed, excited states well above proton-emission threshold do not follow this rule.
The precise calculation of Coulomb contributions for proton drip-line nuclei thus demands the inclusion of continuum coupling in general.
GSM is thus a better tool for that purpose than HO-SM.

\section {Elastic scattering cross sections of the $^{18}$Ne(p,p) reaction}
\begin{figure*}[!htb]
\centering
\includegraphics[width=13cm]{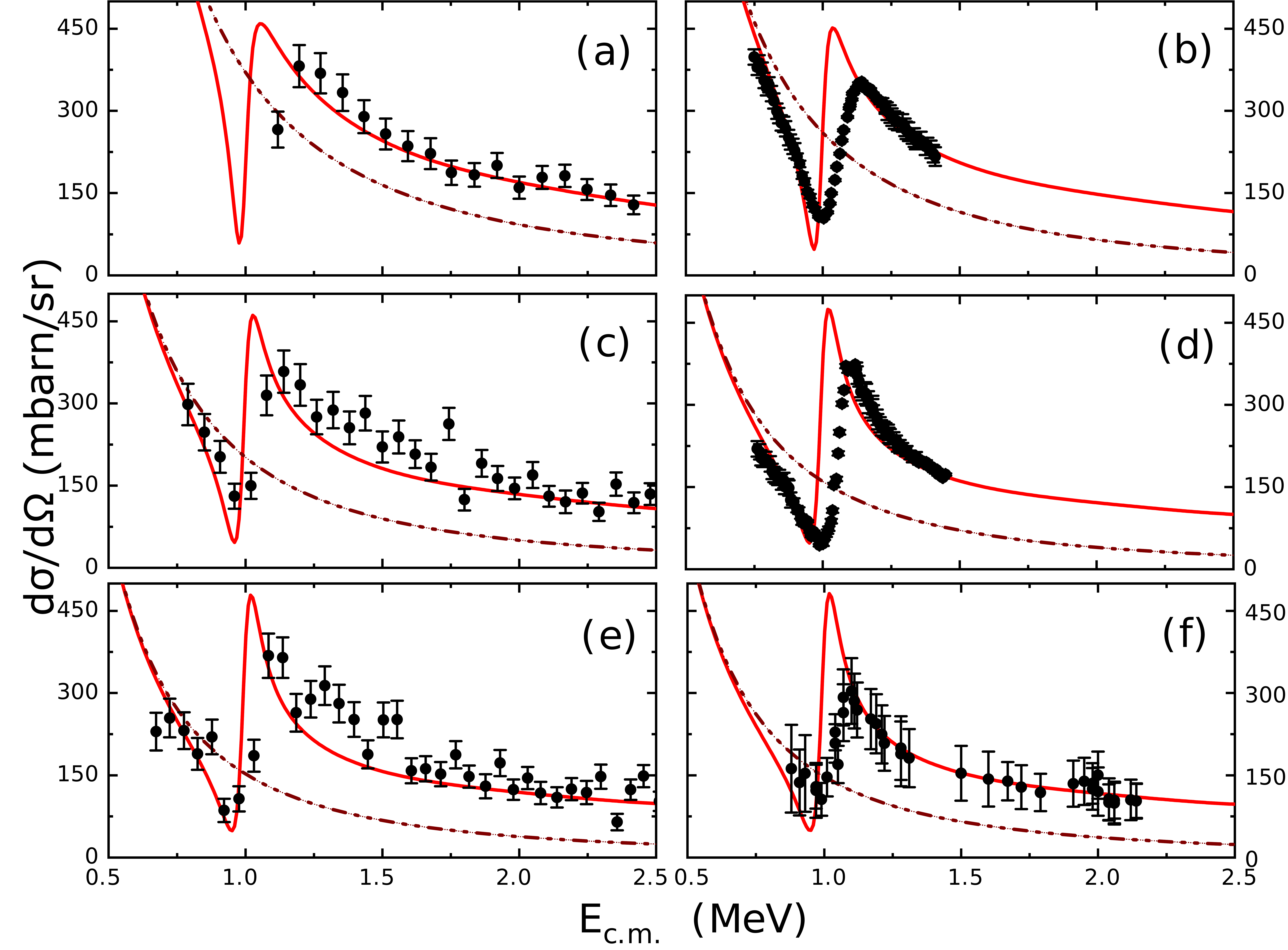}
\caption{Excitation functions of the $^{18}$Ne(p,p) reaction, denoted as $d \sigma / d \Omega$ (mbarn/sr), as a function of the proton projectile center of mass energy $E_{cm}$ (in MeV).
  The panels (a), (b), (c), (d), (e) and (f) respectively depict excitation functions considered at an center of mass angle equal to 105, 120.2, 135, 156.6, 165, and 180 degrees.
  Solid lines correspond to the results obtained with GSM-RGM using the $A$-dependent EFT interaction, dashed lines represent pure Coulomb scattering excitation functions,
  and experimental data, taken from Refs. \cite{PhysRevC.67.014308,deOliveiraSantos2005}, are shown with filled circles (color online).}
\label{scattering}
\end{figure*}
In order to test the predictive power of our approach, we calculated a reaction observable, which is not included in the fitting procedure.
The calculation of reaction observables is indeed meaningful for experimental studies, to better understand nuclear astrophysics experimental data, for example \cite{Thompson_Nunes}.
GSM has thus been supplemented recently by the resonating group method (RGM) \cite{PhysRevC.89.034624}, so that cross sections can be calculated therein.
The most recent applications concern deuteron elastic scattering \cite{PhysRevC.99.044606}, as well as radiative capture \cite{PhysRevC.91.034609,JPG_GX_Dong}.
The excitation functions arising from the $^{18}$Ne(p,p) reaction are of significant interest, as experimental data have been measured in several low-lying energies.
Moreover, these excitation functions depend on the unbound spectrum of $^{19}$Na, so that structure degrees of freedom must be properly included in order to precisely reproduce experimental data.
The excitation functions of the $^{18}$Ne(p,p) reaction had already been calculated in GSM-RGM \cite{PhysRevC.89.034624}.
However, a phenomenological Hamiltonian, fitted on the binding energy and spectra of $^{18}$Ne and $^{19}$Na only, had been used therein.
Indeed, the aim of Ref.\cite{PhysRevC.89.034624} was mainly to demonstrate the feasibility of GSM-RGM calculations.
The situation is thus different, as the used GSM Hamiltonian is fitted on $^{16}$O isotones up to $^{22}$Si,
so that the quality of excitation functions is a direct measure of the ability of the EFT interaction to effectively reproduce experimental data.
The excitation functions of the $^{18}$Ne(p,p) reaction obtained from GSM-RGM are shown in Fig.(\ref{scattering}).
One can see that experimental data are well described. The sudden increase of the excitation function, due to the $1/2^+_1$ excited state of $^{19}$Na, is in particular well reproduced.
However, one can notice a slight shift of the calculated excitation functions compared to experimental data if the angle is equal to 120.2 or 156.6 degrees (see Fig.(\ref{scattering})).
In fact, it would have been necessary to fit the spectra of $^{18}$Ne and $^{19}$Na up to a few keV to match experimental data almost exactly, as was done in Ref.\cite{PhysRevC.89.034624}.
The experimental data of the considered excitation functions are then properly reproduced for all projectile energies.
Consequently, GSM can make efficient predictions of reaction observables starting from a Hamiltonian fitted from structure only.

\section {Conclusion}

The need for continuum coupling might seem less important for the study of nuclei at proton drip-line than for neutron-rich nuclei.
Indeed, due to the presence of the Coulomb barrier, which confines protons in the nuclear region, the wave functions of proton-rich nuclei are expected to resemble those of closed quantum systems.
Moreover, the unbound states of proton-rich nuclei in the medium and heavy regions of the nuclear chart bear life-times of the order of milliseconds, so that continuum effects are small therein.
However, light proton-rich nuclei can exhibit phenomena usually arising only at neutron drip-line, such as halos and resonance character at ground-state level.

The $^{16}$O isotones are particularly interesting for that matter, as the Coulomb barrier is hereby rather important, while peculiarities associated with a sizable continuum coupling still occur.
Hence, $^{16}$O isotones have been studied using GSM, so that the importance of continuum coupling could be precisely assessed,
while the complex nature of inter-nucleon correlations has been handled via the use of two different effective two-body interactions in the presence or absence of $A$-dependence.
The most striking feature seen in $^{16}$O isotones is the importance of three-body forces, or $A$-dependence of the used two-body interaction, equivalently.
While a phenomenological two-body force is sufficient to describe oxygen isotopes at neutron drip-line, leading to an error of at most 500 keV,
a discrepancy of several MeVs has been found to occur in $^{16}$O isotones compared to experimental data if the effects of three-body forces are neglected.
As the total two-body matrix elements of the Hamiltonian are the sum of nuclear and Coulomb parts,
it is reasonable to assume that they are typically smaller in absolute value than in the neutron-rich region, where the nuclear interaction dominates.
As a consequence, the relative importance of three-body forces is larger.
In fact, either a very precise equivalent two-body operator, arising from the normal ordering method \cite{PhysRevC.76.034302},
or an explicit use of $A$-dependence, as done in this work, could provide with a good reproduction of the three-body force effects.
Moreover, while the FHT interaction could reproduce ground states fairly well, the spectra obtained therein were not satisfactory, contrary to the EFT interaction. 
The closer relation of the EFT interaction to realistic nucleon-nucleon interaction might be the reason for that state of affairs, as the FHT interaction is local and, hence, more phenomenological than the EFT interaction.
Indeed, as noted earlier, the partial cancellation of nuclear and Coulomb parts in matrix elements might also enhance the non-local effects of the nuclear interaction,
thus preventing from a good description of $^{16}$O isotones with a local interaction, contrary to neutron-rich oxygen isotopes.
 
The importance of continuum coupling in $^{16}$O isotones could be quantitatively studied.
While the ground states of $^{16}$O isotones are mainly localized in the nuclear region, even when they are unbound, the proton densities of their excited states bear sizable asymptotes.
This generated an interesting pattern in Coulomb contributions, as those associated to the ground states of $^{16}$O isotones follow the IMME prescription,
while those of unbound excited states depart from those values by a few MeVs due to continuum coupling.
Consequently, continuum coupling has to be included if one aims at studying proton-rich nuclei in the $A \sim 20$ region, even if the Coulomb barrier is rather strong therein.
While the ground states of $^{16}$O isotones can be satisfactorily described by closed quantum systems, even when they are of resonance character,
this is not the case for their excited states. Only a proper inclusion of continuum coupling can allow to discriminate between localized and extended many-body states at proton drip-line.

The predictive character of GSM has been tested by considering the excitation function of the $^{18}$Ne(p,p) reaction.
The Hamiltonian has been fitted only from the structure properties of the spectra of $^{16}$O isotones.
The calculated excitation functions of the $^{18}$Ne(p,p) reaction compare well with experimental data for all considered angles.
It is, in particular, due to the good description of the $1/2^+_1$ excited state of $^{19}$Na with GSM, responsible from the fast increase of excitation functions close to $E_{cm}$ = 1 MeV.

\section{Acknowledgments}
This work has been supported by the National Natural Science Foundation of China under Grants No.~11435014, No.~11835001, No.~11921006, No.~11847203 and No.~11975282;
the National Key R\&D Program of China under Grant No.~2018YFA0404401; 
and the CUSTIPEN (China-U.S.~Theory Institute for Physics with Exotic Nuclei) funded by the U.S.~Department of Energy, Office of Science under Grant No.~de-sc0009971.
We acknowledge the High-Performance Computing Platform of Peking University for providing computational resources.

\bibliography{references}

\end{document}